\documentclass[runningheads,a4paper]{llncs}

\usepackage{amssymb}
\usepackage{graphicx}

\usepackage{url}
\urldef{\mailsa}\path|muhammad.khan@dk-compmath.jku.at|
\urldef{\mailsb}\path|Wolfgang.Schreiner@risc.jku.at|

\begin{document}

\mainmatter  

\title{On Formal Specification of Maple Programs\thanks{The research was funded by the Austrian Science Fund (FWF): W1214-N15, project DK10.}}

\titlerunning{On Formal Specification of Maple Programs}

%
%
\author{Muhammad Taimoor Khan%
\and Wolfgang Schreiner}
\authorrunning{Muhammad Taimoor Khan and Wolfgang Schreiner}

\institute{Doktoratskolleg Computational Mathematics\\
and\\
Research Institute for Symbolic Computation\\
Johannes Kepler University\\
Linz, Austria\\
\mailsa\\
\mailsb\\
\url{http://www.risc.jku.at/people/mtkhan/dk10/}}

%
%

\maketitle

\enlargethispage*{1.5cm}
\begin{abstract}
This paper is an example-based demonstration of our initial results on the formal specification of programs written in the computer algebra language \emph{MiniMaple} (a substantial subset of Maple with slight extensions). The main goal of this work is to define a verification framework for \emph{MiniMaple}.
Formal specification of \emph{MiniMaple} programs is rather complex task as it supports non-standard types of objects, e.g. symbols and unevaluated expressions, and additional functions and predicates, e.g. runtime type tests etc.
We have used the specification language to specify various computer algebra concepts respective objects of the Maple package \emph{Difference\-Diff\-eren\-tial} developed at our institute. 
\end{abstract}

\section{Introduction}
We report on a project whose goal is to design and develop a tool to find behavioral errors such as type inconsistencies and violations of method preconditions in programs written in the language of the computer algebra system Maple; for this purpose, these programs need to be annotated with the types of variables and methods contracts~\cite{Meyer1992}.

As a starting point, we have defined a substantial subset of the computer algebra language Maple, which we call \emph{MiniMaple}. Since type safety is a prerequisite of program correctness, we have formalized a type system for \emph{MiniMaple} and implemented a corresponding type checker. The type checker has been applied to the Maple package \emph{DifferenceDifferential}~\cite{ChrDon09} developed at our institute for the computation of bivariate difference-differential dimension polynomials. Furthermore, we have defined a language to formally specify the behavior of \emph{MiniMaple} programs. As the next step, we will develop a verification calculus for \emph{MiniMaple}. The other related technical details about the work presented in this paper are discussed in the accompanying paper~\cite{MTK12c}. For project details and related software, please visit \url{http://www.risc.jku.at/people/mtkhan/dk10/}.

The rest of the paper is organized as follows: in Section 2, we briefly demonstrate formal type system for \emph{MiniMaple} by an example. In Section 3, we introduce and demonstrate the specification language for \emph{MiniMaple} by an example. Section 4 presents conclusions and future work.

\enlargethispage*{2cm}
\section{A Type System for \emph{MiniMaple}}

\emph{MiniMaple} procedure parameters, return types and corresponding local (variable) declarations needs to be (manually) type annotated. Type inference would be partially possible and is planed as a later goal. The results we derive with type checking Maple can also be applied to Mathematica, as Mathematica has almost the same kinds of runtime objects as Maple.

Listing 1 gives an example of a \emph{MiniMaple} program which we will use in the following section for the discussion of type checking respective formal specification. Also the type information produced by the type system is shown by the mapping $\pi$ of program variables to types. For other related technical details of the type system, please see~\cite{MTK11a}.
\begin{footnotesize}
\begin{tabbing} \emph{1}. status:=0; \\ \emph{2}. prod := \= \textbf{proc}(l::\textbf{list}(\textbf{Or}(\textbf{integer},\textbf{float})))::[\textbf{integer},\textbf{float}];
\\\emph{3}. \> \textsl{\# $\pi$=\{l:\textbf{list}(\textbf{Or}(\textbf{integer},\textbf{float}))\}}
\\\emph{4}. \> \textbf{global} status; 
\\\emph{5}. \> \textbf{local} i, x::\textbf{Or}(\textbf{integer},\textbf{float}), si::\textbf{integer}:=1, sf::\textbf{float}:=1.0;
\\\emph{6}. \> \textsl{\# $\pi$=\{..., i:\textbf{symbol}, x:\textbf{Or}(\textbf{integer},\textbf{float}),..., status:\textbf{anything}\}}
\\\emph{7}. \> \textbf{for} \= i \textbf{from} 1 \textbf{by} 1 \textbf{to} nops(l) \textbf{do}
\\\emph{8}. \> \> x:=l[i]; status:=i;
\\\emph{9}. \> \> \textsl{\# $\pi$=\{..., i:\textbf{integer}, ..., status:\textbf{integer}\}}
\\\emph{10}. \> \> \textbf{if} \textbf{type}(\= x,integer) \textbf{then}
\\\emph{11}. \> \> \textsl{\# $\pi$=\{..., i:\textbf{integer}, x:\textbf{integer}, si:\textbf{integer}, ..., status:\textbf{integer}\}}
\\\emph{12}. \> \> \> \textbf{if} (\=x = 0) \textbf{then}
\\\emph{13}. \> \> \> \> \textbf{return} [si,sf];
\\\emph{14}. \> \> \> \textbf{end if};
\\\emph{15}. \> \> \> si:=si*x;
\\\emph{16}. \> \> \textbf{elif} \textbf{type\=}(x,float) \textbf{then}
\\\emph{17}. \> \> \textsl{\# $\pi$=\{..., i:\textbf{integer}, x:\textbf{float}, ..., sf:\textbf{float}, status:\textbf{integer}\}}
\\\emph{18}. \> \> \> \textbf{if} (\=x $<$ 0.5) \textbf{then}
\\\emph{19}. \> \> \> \> \textbf{return} [si,sf];
\\\emph{20}. \> \> \> \textbf{end if};
\\\emph{21}. \> \> \> sf:=sf*x;
\\\emph{22}. \> \>\textbf{end if};
\\\emph{23}. \> \> \textsl{\# $\pi$=\{..., i:\textbf{integer}, x:\textbf{Or}(\textbf{integer},\textbf{float}),..., status:\textbf{integer}\}}
\\\emph{24}. \> \textbf{end do;}
\\\emph{25}. \> \textsl{\# $\pi$=\{..., i:\textbf{symbol}, x:\textbf{Or}(\textbf{integer},\textbf{float}),..., status:\textbf{anything}\}}
\\\emph{26}. \> status:=-1;
\\\emph{27}. \> \textsl{\# $\pi$=\{..., i:\textbf{symbol}, x:\textbf{Or}(\textbf{integer},\textbf{float}),..., status:\textbf{integer}\}}
\\\emph{28}. \> \textbf{return} [si,sf];
\\\emph{29}. \> \textbf{end proc};
\\\emph{30}. result := prod([1, 8.54, 34.4, 6, 8.1, 10, 12, 5.4]);
\end{tabbing}
\end{footnotesize}
\begin{center}
 Listing 1: The example \emph{MiniMaple} procedure type-checked
\end{center}

\enlargethispage*{2.5cm}
The following problems arise from type checking \emph{MiniMaple} programs:
\begin{itemize}
 \item Global variables (declarations) can not be type annotated; therefore values of arbitrary types can be assigned to global variables in Maple. Therefore we introduce \emph{global} and \emph{local} contexts to handle the different semantics of the variables inside and outside of the body of a procedure respectively loop.
 \begin{itemize}
  \item In a \emph{global} context new variables may be introduced by assignments and the types of variables may change arbitrarily by assignments.
  \item In a \emph{local} context variables can only be introduced by declarations. The types of variables can only be \emph{specialized} i.e. the new value of a variable should be a sub-type of the declared variable type. The sub-typing relation is observed while specializing the types of variables.
 \end{itemize}
  \item A predicate \textbf{type}(\emph{E,T}) (which is true if the value of expression $E$ has type $T$) may direct the control flow of a program. If this predicate is used in a conditional, then different branches of the conditional may have different type information for the same variable. We keep track of the type information introduced by the different type tests from different branches to adequately reason about the possible types of a variable. For instance, if a variable x has type Or(integer,float), in a conditional statement where the "if" branch is guarded by a test 
   type(x,integer), in the "else" branch x has automatically type float. This automatic type inferencing only applies if an identifier has a union type. A warning is generated, if a test is redundant (always yields true or false).

\end{itemize}

The type checker has been applied to the Maple package \emph{Difference\-Differential}~\cite{ChrDon09}. No crucial typing errors have been found but some bad code parts have been identified that can cause problems, e.g., variables that are declared but not used (and therefore cannot be type checked) and variables that have duplicate global and local declarations.

\enlargethispage*{2.2cm}
\section{A Specification Language for \emph{MiniMaple}}
Based on the type system presented in the previous section, we have developed a formal specification language for \emph{MiniMaple}. This language is a logical formula language which is based on Maple notations but extended by new concepts. The formula language supports various forms of quantifiers, logical quantifiers (\textbf{exists} and \textbf{forall}), numerical quantifiers (\textbf{add}, \textbf{mul}, \textbf{min} and \textbf{max}) and sequential quantifier (\textbf{seq}) representing truth values, numeric values and sequence of values respectively. We have extended the corresponding Maple syntax, e.g., logical quantifiers use typed variables and numerical quantifiers are equipped with logical conditions that filter values from the specified variable range.

Also the language allows to formally specify the behavior of procedures by pre- and post-conditions and other constraints; it also supports loop specifications and assertions. In contrast to specification languages such as Java Modeling Language~\cite{JML}, abstract data types can be introduced to specify abstract concepts and notions from computer algebra.



\begin{footnotesize}
\hspace*{-6 mm}(*@
\\ \textbf{requires} true;
\\ \textbf{global} status;
\\ \textbf{ensures}
\\ \hspace*{3 mm}(status = -1 and RESULT[1] = mul(e, e in l, type(e,integer))
\\ \hspace*{3 mm}and RESULT[2] = mul(e, e in l, type(e,float))
\\ \hspace*{3 mm}and forall(i::integer, 1$<=$i and i$<=$nops(l) and type(l[i],integer) implies l[i]$<>$0)
\\ \hspace*{3 mm}and forall(i::integer, 1$<=$i and i$<=$nops(l) and type(l[i],float) implies l[i]$>=$0.5))
\\ \hspace*{3 mm}or
\\ \hspace*{3 mm}(1$<=$status and status$<=$nops(l) and RESULT[1] = mul(l[i], i=1..status-1, type(l[i],integer))
\\ \hspace*{3 mm}and RESULT[2] = mul(l[i], i=1..status-1, type(l[i],float))
\\ \hspace*{3 mm}and ((type(l[status],integer) and l[status]=0) or (type(l[status],float) and l[status]$<$0.5))
\\ \hspace*{3 mm}and forall(i::integer, 1$<=$i and i$<$status and type(l[i],integer) implies l[i]$<>$0)
\\ \hspace*{3 mm}and forall(i::integer, 1$<=$i and i$<$status and type(l[i],float) implies l[i]$>=$0.5));
\\ @*)
\\ \textbf{proc}(l::\textbf{list}(\textbf{Or}(\textbf{integer},\textbf{float})))::[\textbf{integer},\textbf{float}]; ... \textbf{end proc};
\end{footnotesize}
\begin{center}
 Listing 2: The example \emph{MiniMaple} procedure formally specified
\end{center}

Listing 2 gives a formal specification of the example procedure introduced in Section 2. The procedure has no pre-condition as shown in the \textbf{requires} clause; the \textbf{global} clause says that a global variable \emph{status} can be modified by the body of the procedure. The normal behavior of the procedure is specified in the \textbf{ensures} clause. The post condition specifies that, if the complete list is processed then we get the result as the product of all integers and floats in the list but if procedure terminates pre-maturely then we only get the product of integers and floats till the value of variable \emph{status} (index of the input list). For the complete syntax and other details of the formal specification language see~\cite{MTK11c}. To test the specification language, we have formally specified some parts of the Maple package \emph{DifferenceDifferential}~\cite{ChrDon09} developed at our institute as the main test for the specification language.

\enlargethispage*{2cm}
\section{Conclusions}

We may use the specification language sketched in this short paper to generate executable assertions that are embedded in \emph{Mini\-Maple} programs and check at runtime the validity of pre/post conditions. Our main goal, however, is to use the specification language to verify the correctness of \emph{MiniMaple} annotated programs by static analysis, in particular to detect violations of methods preconditions. For this purpose, based on the results of a prior investigation, we intend to use the verification framework Why3~\cite{Why3} to implement the verification calculus for \emph{MiniMaple}, i.e., to translate \emph{MiniMaple} into the intermediate language of Why3 and to apply its verification condition generator to generate verification conditions and prove their correctness with various back-end provers. Since the verification calculus must be sound, we have defined a formal semantics of \emph{MiniMaple}~\cite{MTK12a} such that the correctness of the transformation can be shown. 



\enlargethispage*{2cm}
\begin{small}
\bibliography{paper}

\begin{thebibliography}{1}

\bibitem{Why3}
Fran\c{c}ois Bobot, Jean-Christophe Filli\^atre, Claude March\'e, and Andrei
  Paskevich.
\newblock Why3: Shepherd your herd of provers.
\newblock In {\em Boogie 2011: First International Workshop on Intermediate
  Verification Languages}, Wroc\l{}aw, Poland, August 2011.

\bibitem{ChrDon09}
{{{Christian D{\"o}nch}}}.
\newblock {{Bivariate Difference-Differential Dimension Polynomials and Their
  Computation in Maple}}.
\newblock Technical report, Research Institute for Symbolic Computation,
  Johannes Kepler University, Linz, 2009.

\bibitem{JML}
{Gary T. Leavens and Yoonsik Cheon}.
\newblock {Design by Contract with JML}.
\newblock {A Tutorial}, {2006}.
\newblock \url{ftp://ftp.cs.iastate.edu/pub/leavens/JML/jmldbc.pdf}.

\bibitem{MTK11a}
Muhammad~Taimoor Khan.
\newblock {A Type Checker for \emph{MiniMaple}}.
\newblock {RISC Technical Report} 11-05, also DK Technical Report 2011-05,
  Research Institute for Symbolic Computation, Johannes Kepler University,
  Linz, 2011.

\bibitem{MTK12a}
Muhammad~Taimoor Khan.
\newblock {Formal Semantics of \emph{MiniMaple}}.
\newblock {DK Technical Report} 2012-01, Research Institute for Symbolic
  Computation, Johannes Kepler University, Linz, January 2012.

\bibitem{MTK11c}
Muhammad~Taimoor Khan and Wolfgang Schreiner.
\newblock {Towards a Behavioral Analysis of Computer Algebra Programs (Extended
  Abstract)}.
\newblock In Paul Pettersson and Cristina Seceleanu, editors, {\em {Proceedings
  of the 23rd Nordic Workshop on Programming Theory (NWPT'11)}}, pages 42--44,
  Vasteras, Sweden, October 2011.

\bibitem{MTK12c}
Muhammad~Taimoor Khan and Wolfgang Schreiner.
\newblock {{Towards the Formal Specification and Verification of Maple
  Programs}}.
\newblock In {\em {Conferences on Intelligent Computer Mathematics, Calculemus
  track (submitted)}}. 2012.

\bibitem{Meyer1992}
Bertrand Meyer.
\newblock {Applying Design by Contract}.
\newblock {\em Computer}, 25:40--51, October 1992.

\end{thebibliography}
\bibliographystyle{plain}
\end{small}
\end{document}